# The value of information in financial markets: An agent-based simulation

Bence Tóth[1,2] and Enrico Scalas[3]

[1] ISI Foundation - Torino, Italy
[2] Budapest University of Technology and Economics – Budapest, Hungary
[3] Dipartimento di Scienze e Tecnologie Avanzate, East Piedmont University - Alessandria, Italy

## Abstract

We present results on simulations of a stock market with heterogeneous, cumulative information setup. We find a non-monotonic behaviour of traders' returns as a function of their information level. Particularly, the average informed agents underperform random traders; only the most informed agents are able to beat the market. We also study the effect of a strategy updating mechanism, when traders have the possibility of using other pieces of information than the fundamental value. These results corroborate the latter ones: it is only for the most informed player that it is rewarding to stay fundamentalist. The simulations reproduce some stylized facts of tick-by-tick stock-exchange data and globally show informational efficiency.

## Motivation: Experiments and computer modeling

In many studies on the role/value of information in markets, including stock markets, usually two levels of information, such as informed and uninformed, are taken into account (Grossman and Stiglitz 1980, Hellwig 1982, Figliewski 1982, Sunder 1992, Scalas et al. 2005). In this case, more knowledge is always better then less. In other words, insiders are able to outperform uninformed investors. However, interesting behaviour may emerge if one takes into account the interaction of more than two persons with heterogeneous information levels. There is a physical analogy when considering two-body systems versus many-body systems. Many-body systems display complex behaviour such as chaotic sensitivity to initial conditions. Many heterogeneous agents in a stock market may interact both directly, by direct communication, and indirectly, via the market mechanism. In this paper, we model a set of experiments where more than two agents endowed with heterogeneous information trade a single risky asset using a double auction mechanism and do not communicate among each other. The agents can choose between the risky asset paying a random dividend and cash for which a fixed interest rate is paid at the end of each trading period.

Experiments based on the previous-mentioned ideas were carried out by Jürgen Huber and Michael Kirchler at the University of Innsbruck. In particular, we focused on the 2004 session with the participation of business students[*] (Huber et al. 2006, Kirchler and Huber 2007). These experiments were based on a cumulative information system: More informed agents had additional information compared to less informed agents. Information was defined by the forecasting horizon of future dividends (which were generated before the trading session)[**]. Nine traders with different forecasting abilities were trading in a continuous double auction with limit orders and market orders. As mentioned above, at the end of each period, a risk free interest rate was paid on the cash held by the traders and dividends were paid based on the shares owned, with parameters set so that one period corresponds to one month in real market.

The market and information setup leads to the following educated guess on the performance of traders as a function of information: Since the model can be understood as a zero sum game compared to the market return, we certainly know, that if an uninformed (random) trader earns the market average return, then either all traders get exactly the market return or, if the less informed agent approximately gets the market return, performance will be a non monotonic function of the information level (Schredelseker 1984).

The experiments gave interesting results (Huber 2006, Huber et al. 2006). The net return of traders compared to the market return as a function of the information level does not show a monotonic growth with increasing information. Traders having the first five levels of information do not outperform the average and only the best-informed traders (insiders) were able to get excess return.

The authors also analyzed some well-known empirical facts (the so-called stylized facts) of real world markets. The probability density function of price changes, the decay of the autocorrelation function of price changes and the decay of the autocorrelation function of absolute price changes were studied. A further study on durations was also performed. In all the cases, the experimental market was able to reproduce the stylized facts: the distribution of returns was fat tailed, the autocorrelation of returns decayed fast, the autocorrelation of absolute returns decayed slowly (volatility clustering), and the durations were not exponentially distributed. (Kirchler and Huber 2007, Scalas et al. 2006).

## The simulations

### The market

In our simulation, we programmed an essential double auction trading mechanism as it appears on most of real world financial markets, with a book containing bid and ask orders. As, contrary to real world experiments, in a numerical simulation one has the possibility to implement truly random traders, we simulated ten agents with different levels of information going from zero (random traders), *I0* to nine *I9,* and with the possibility of using different trading strategies as will be discussed in details in the following subsections. As in the experiments, these levels coincided with the forecasting horizon of each trader.

---

[*] The experiments and their results are described in details in other papers of the present book.
[**] The market mechanism and the information setup will be described in details later in the section The simulations.

In order to be able to estimate the statistical error in our results, we carried out 10000 runs in each simulation (100 sessions of 100 runs).

The market contained a risky asset (stock) and a risk free bond (cash). Before beginning the simulation an information level was assigned to each of the ten agents (nine informed and one uninformed), thus having one agent for every level. Initially all the agents were given 1600 units of cash and 40 shares of stock with initial value of 40 units each. Trading consisted of 30 periods each lasting 100 simulation steps. At the beginning of each period new information was delivered to the agents according to their information level.

At the end of each period a risk free interest rate was paid on the cash held by the agents, dividends were paid on the shares held by the agents (the risk free interest rate was $r_f=0.01$, the risk adjusted interest rate $r_e=0.005^*$) and the book was cleared. We also carried out simulations without clearing the book and found that the clearing process does not make much difference in the results.

The dividend process (being the source of future information) was determined before the beginning of trading, being a random walk of Gaussian steps:

$$D(i) = D(i-1) + 0.1 N(0,1),$$

with $D(0)=0.2$, where $N(0,1)$ is a normal distribution with zero mean and unit variance (in case of $D(i)<0$, we took $|D(i)|$). We carry out finite time simulations, so that short trends in the random walk can have important effects on the dividend process and by that on the information structure and the price formation on the market. When studying the performance of heterogeneously informed agents, we used different dividend processes. The results shown are the statistics of 100 simulation sessions, each being run with a different random dividend process. One session consists of 100 independent simulation runs carried out with the same dividend process.

**Information**

In order to value shares, traders on the market got information about future dividends. The idea of Hellwig 1982 was extended to ten information levels: different levels of information correspond to different lengths of windows in which one can predict future dividends. Trader *I0* is completely uninformed and put orders at random, trader *I1* knows the dividend at the end of the current period, trader *I2* knows the dividends at the current and the next period, and so on up to trader *I9* knowing the current and the next eight period dividends (Huber 2006, Huber et al. 2006). In this way, we got a cumulative information structure of the market where better-informed agents know future dividends earlier than less informed ones. Since market trading consists of several periods (new information entering the market in each), the design implies that information trickles down through the market from the more informed to the less informed over time.

The information obtained by traders is the present value of the stock conditioned on the forecasting horizon ($E(V|I_{j,k})$). This is calculated using Gordon's formula, discounting the known dividends and assuming the last one as an infinite stream, which is also

---

[*] In the paper Tóth et al. 2007 the risk adjusted interest rate was erroneously given as $r_e=0.05$.

discounted. E(V|I_{j,k}) stands for the conditional present value of the asset in period *k* for the trader with information level *j* (*Ij*).

$$E(V \mid I_{j,k}) = \frac{D(k+j-1)}{r_e(1+r_e)^{j-2}} + \sum_{i=k}^{k+j-2} \frac{D(i)}{(1+r_e)^{i-k}},$$

where $r_e$ is the risk adjusted interest rate (E(-- | --) denotes the conditional average).

Before the beginning of the experiment, an information level was randomly assigned to each agent. There was one trader for each information level. At the beginning of each period new information was delivered to the agents depending on their level of information.

Summarizing, as often mentioned, overall we implemented ten levels of information, a completely uninformed trader (random trader), *I0* and nine informed traders with different levels of information from *I1* to *I9*, where agent *Ij* has information of the dividends for the end of the current period and of (j-1) forthcoming periods (forecasting ability).

**Trading strategies**

At the beginning of each period, agents submit orders according to their idea of the value of stocks. After that, during the period, in every time step, a trader is randomly chosen, who either accepts a limit order from the book (puts a market order) or puts a new limit order to the book.

Since we do not exactly know how traders use their information in the real world and in the experiments, we gave the possibility to simulated traders to strictly apply the fundamental information they get (*fundamentalists*), not to take any information in account except the current price and trade randomly (*random traders*) or to look at other pieces of information such as trends (*chartists*). In this section we describe the different strategies in short. The Mat Lab® codes for these strategies can be found in Appendix A, as well as on the web page: http://www.phy.bme.hu/~bence.

*Fundamentalists*

Fundamentalist traders strictly believe in the information they receive. If they find an *ask order* with a price lower or a *bid order* with a price higher than their estimated present value, i.e. $E(V|I_{j,k})$, they accept the limit order, otherwise they put a new limit order between the former best bid and best ask prices.

*Random traders*

Random traders randomly send orders to the market based on the price of the last trade (current price). With probability 0.5 they put an ask (bid) order slightly higher (lower) than the current price.

*Chartists*

Chartist traders are searching for trends in the price movements. In case of three consecutive upward (downward) price steps they buy (sell), otherwise they give a new order to the limit order book.

# Results

In our simulations we considered the effect of information on the performance of agents throughout the market session and we studied the returns of different traders as a function of the information level. We also analysed the results from the point of view of the Efficient Market Hypothesis and we qualitatively checked the stylized facts of stock markets. As already mentioned, in order to reduce statistical errors we carried out 10000 runs of the simulation. Finally, when thinking about real world situations, one has to take into account some kind of learning/adaptation process for the traders (Schredelseker 2001). To mimic this effect, we also studied the situation when agents can periodically compare their performance to the market average and change their trading strategies if they find themselves under the average.

**Final wealth as a function of information**

The final return of each agent relative to the market return can be seen in Figure 1, the results are the average of 100 sessions, each session consisting of 100 runs. These results are in good qualitative agreement with the experimental results presented elsewhere in this book: we get a curve that we call *J-curve*. Agents with an average level of information (*I1-I5*) perform worse than the completely uninformed random agent (*I0*). The most informed agents outperform the market. To test the hypothesis of the J-curve we ran the Wilcoxon rank sum test for equal medians (Gibbons 1985, Hollander and Wolfe 1973), on the relative return for pairs of information levels. The p-values of the tests can be found in Table 1. One can see that the hypothesis of two returns corresponding to different agents being drawn from the same distribution can be excluded in almost all cases at the 0.05 significance level. This result and its relevance for real markets will be discussed further in the section Conclusions.

|    | *I0*    | *I1*    | *I2*    | *I3*    | *I4*    | *I5*    | *I6*    | *I7*    | *I8* |
|----|---------|---------|---------|---------|---------|---------|---------|---------|------|
| *I1* | 0.000*  |         |         |         |         |         |         |         |      |
| *I2* | 0.000*  | 0.932   |         |         |         |         |         |         |      |
| *I3* | 0.000*  | 0.507   | 0.385   |         |         |         |         |         |      |
| *I4* | 0.000*  | 0.003*  | 0.002*  | 0.013*  |         |         |         |         |      |
| *I5* | 0.000*  | 0.000*  | 0.000*  | 0.000*  | 0.000*  |         |         |         |      |
| *I6* | 0.144   | 0.000*  | 0.000*  | 0.000*  | 0.000*  | 0.000*  |         |         |      |
| *I7* | 0.000*  | 0.000*  | 0.000*  | 0.000*  | 0.000*  | 0.000*  | 0.000*  |         |      |
| *I8* | 0.000*  | 0.000*  | 0.000*  | 0.000*  | 0.000*  | 0.000*  | 0.000*  | 0.009*  |      |

| I9 | 0.000* | 0.000* | 0.000* | 0.000* | 0.000* | 0.000* | 0.000* | 0.001* | 0.057** |

**Table 1 p-values of the Wilcoxon rank sum test for equal medians on differences in performance between the information levels.**

**\* significant at the $0.05$ level**

**\** significant at the $0.1$ level**

In order to understand why the random trader gets almost exactly the market return and to see how the relative wealth of agents looks like in simpler cases, we ran simulations with only three agents in the market (the standard deviation of dividends being 0.01, $r_f$=0.001 and $r_e$=0.005 in this case): an uninformed (*I0*), an average informed (*I4*) and a well informed (*I9*).

Figure 2 shows the same plot of Figure 1 for these simulations with only three agents. Again, we can exclude the monotonicity of the curve and even if with three points it is harder to call it a J-curve, we can see that the random trader performs better than the average informed one and only the well-informed agent gets excess returns. In this case also the performance of the random trader is below the market level, giving an explanation based on price impact for the question raised: when enough actors are present on the market, the price impact of the random trader becomes negligible, thus the random trader has equal probability of being beaten by the market and of beating the market.

In order to gain more insight into the process, in Table 2 we show the relative performance of the random trader in the case of different number of agents (when the least informed agents are always present on market).

| number of traders | 3 | 5 | 7 | 9 | 10 |
|---|---|---|---|---|---|
| relative performance of random trader [%] | -2.7 | -1.0 | -0.2 | +0.2 | -0.1 |

**Table 2 Relative performance of the random trader on markets with different number of agents. Note that in this case the least informed agents are always present on the market (e.g. in case of 3 traders: *I0*, *I1* and *I2*). As more agents are present on the market, the performance of the random agent approaches 0 (the market return).**

**The Efficient Market Hypothesis**

Efficiency has various meanings in economics. One can talk about *productive efficiency*, *Pareto efficiency* or *informational efficiency* (Tesfatsion 2006). When we write the term efficiency, we mean informational efficiency: A market is efficient if all information available is built into the asset prices at each time instance. In other words, the price of an asset reflects all available relevant information. Thus a market containing risky financial assets is efficient in a time period if the traders exploit all profit opportunities, with no further profit to be achieved. In case of risk-neutral traders, for a market containing a risk-free asset, this leads to the following sub-martingale condition: the conditional expected return on the risky asset must be equal to the return rate of the risk free asset.

Let the return rate on the risk-free asset in the period [t,t+1) be denoted by $r_f(t,t+1)$, and $r_A(t,t+1)$ be the return for the risky asset *A* in the same period. Let us denote the information available to the traders at time *t* by *I(t)*. Then the efficient market hypothesis can be written in the following way:

$$E[r_A(t,t+1) | I(t)] = r_f(t,t+1), \qquad t=0,1,2,...$$

Where, as above, E(-- | --) denotes the conditional expectation.

We tested our simulations from the point of view of efficiency as described above. In our case, the risk adjusted interest rate, which we used for discounting future dividends was set to $r_e=0.005$. The histogram of the net simple returns on the asset is presented in Figure 3. One can see that the average of net simple returns in the simulations is 0.0049, very close to the adjusted interest rate. Thus, we can state that the simulated market globally shows informational efficiency.

**Stylized facts**

A tool for corroborating the relevance of results in artificial markets is analysing whether the empirical stylized facts of markets (Cont 2001) are present. While not getting stylized facts in a simulation can falsify the assumptions made, the presence of these facts does not confirm the hypotheses as well as other results of the simulation.
We considered three common empirical stylized facts as in Kirchler and Huber 2007 since the synchronous time structure of our simulation is unrealistic (Daniel 2006, Muchnik et al. 2006). Figure 4 shows the autocorrelation functions of returns (circles and lines) and of absolute returns (dots and lines). The noise level of the computations is also included in the plot (straight lines). One can see that the autocorrelation of returns immediately decays under the noise level (with a negative overshoot for small lags as usual for high frequency data in real world markets too - in real world markets the negative overshoot is explained by market friction due to fees, here it is a likely consequence of the unrealistic time structure), thus there are no long range time correlation in signed returns. On the other hand, the autocorrelation of absolute returns slowly decays showing that price fluctuations tend to cluster (volatility clustering). (A slight even-odd oscillation is visible in the autocorrelation of absolute returns; this is an artifact of our simulation process, as in many cases, the inter-trade duration is two simulation steps, resulting in this oscillation.)

For comparison, in Table 3 we have reported the first four moments of the log-return distribution for tick-by-tick General Electric (GE) prices in October 1999 (55559 data points). In Figure 5, for these data, the autocorrelation of signed and absolute log-returns is plotted as in Figure 4. In Table 3 it can be seen that the distribution function of absolute returns in the simulation is leptokurtic, similarly to real world data, even if the kurtosis is lower. Running the Jarque-Bera test, for goodness-of-fit to a normal distribution (Judge et al. 1988), we can rule out the normality of the distribution of the absolute returns for both cases.

|                    | GE data     | simulation  |
|--------------------|-------------|-------------|
| mean               | 2.1215e-06  | 3.5221e-05  |
| standard deviation | 4.0078e-04  | 0.0250      |
| skewness           | -0.0698     | 0.1233      |
| kurtosis           | 36.2677     | 7.9541      |

**Table 3 The first four moments of the logarithmic return distribution for tick-by-tick General Electric (GE) prices and for the simulated prices.**

**The Markov chain for switching strategies**

In real life situations, one cannot expect that a trader who is performing under the market average for an extended period, to stick to his/her trading strategy. There are several possibilities for a trader to revise his/her strategy. He/she can try looking at other pieces of information, changing the stocks he/she is investing in or moving to other sectors in which he/she is interested in, etc.; we tried to mimic this freedom and we implemented a strategy updating method: Periodically all traders look around and compare their return in a certain time interval to the return of the market index in the same interval. If they find that they performed under the market they change strategy: If they were fundamentalists they switch to chartist, if they were chartists they switch to fundamentalist.

An interesting question is, whether there exists a stable set of strategies depending on the fundamental information level (information which is only used while acting as a fundamentalist, but we continue labeling traders according to this level).

We have run extended simulations searching for a stationary distribution in the case of three and five agents corresponding to the levels *I1 – I3* and *I1 – I5,* respectively.

When searching for a stationary distribution, we use notions from Markov chain theory and Monte Carlo sampling. Each agent in each period can be either a fundamentalist or a chartist, defining the states of the system between which there are transitions. The strategy space in our system has $2^n$ states, where n denotes the number of agents.

The $2^n \times 2^n$ transition matrix between the states is defined as:

$$\mathbf{T}(A,B) = \frac{\text{number of transitions } (A \Rightarrow B)}{\text{number of occurences } (A)},$$

where *A, B* are states of the system. Note that since agents compare their performance to the market average, the diagonal and anti-diagonal elements of the transition matrix are zero by definition. The frequency of states is denoted by the vector **π**, defined by:

$$\pi = (\pi(1),...,\pi(2^n)),$$

with

$$\pi(A) = \frac{\text{number of occurences (A)}}{\text{total number of occurences}},$$

so that $\sum_A \pi(A) = 1$.

A distribution of frequencies is stationary if

$$\pi T = \pi,$$

where $\pi$ denotes the transpose of $\pi$.

*3 traders*

In the case of three traders, the system has eight possible states. We ran simulations for 100000 periods with strategy updating at the end of every period. We studied the system with all the eight possible initial conditions (the codes for the states can be found in Appendix B). Figure 6 shows the mean of the frequencies of states over all possible initial states, with the standard deviations shown in the error bars. We find that regardless of the initial state of the system, the frequencies are approximately the same.

As one can see, there are two states (in other words one state pair) with higher frequencies, while the remaining six states appear with much smaller and almost equal frequency. Table 4 shows the transition frequencies between the states of the system. From the transition matrix we can see that the transition frequencies between states 2 and 3 and vice versa are higher, thus justifying the notion of "state pair". Taking a closer look at the most probable states we can see that they correspond to the setups: I1=C, I2=F, I3=F (state 2) and I1=F, I2=C, I3=F (state 3). (Where F stands for fundamentalist, C stands for chartist). In both cases the most informed trader (*I3*) remains fundamentalist (thus for him/her it is always worth using his/her information), while the less informed traders bounce between the two possible strategies. This result points to the same direction as the results discussed above: It is only for the most informed trader that it is worth using his/her information.

| state | 1 | 2 | 3 | 4 | 5 | 6 | 7 | 8 |
|---|---|---|---|---|---|---|---|---|
| 1 | 0±0 | 0.146± 0.0042 | 0.134± 0.0047 | 0.255± 0.0073 | 0.169± 0.0040 | 0.149± 0.0047 | 0.144± 0.0063 | 0±0 |
| 2 | 0.037± 0.0020 | 0±0 | 0.816± 0.0063 | 0.031± 0.0013 | 0.038± 0.0022 | 0.045± 0.0018 | 0±0 | 0.032± 0.0010 |
| 3 | 0.034± 0.0019 | 0.824± 0.0052 | 0±0 | 0.030± 0.0015 | 0.030± 0.0007 | 0±0 | 0.042± 0.0017 | 0.037± 0.0017 |
| 4 | 0.037± 0.0017 | 0.145± 0.0041 | 0.132± 0.0069 | 0±0 | 0±0 | 0.119± 0.0040 | 0.134± 0.0058 | 0.207± 0.0140 |
| 5 | 0.205± 0.0090 | 0.162± 0.0085 | 0.139± 0.0060 | 0±0 | 0±0 | 0.145± 0.0042 | 0.145± 0.0039 | 0.200± 0.0039 |
| 6 | 0.153± 0.0053 | 0.193± 0.0066 | 0±0 | 0.140± 0.0056 | 0.144± 0.0038 | 0±0 | 0.219± 0.0066 | 0.149± 0.0024 |
| 7 | 0.142± 0.0036 | 0±0 | 0.188± 0.0095 | 0.151± 0.0057 | 0.143± 0.0059 | 0.213± 0.0074 | 0±0 | 0.160± 0.0045 |
| 8 | 0±0 | 0.127± 0.0080 | 0.130± 0.0054 | 0.287± 0.0132 | 0.170± 0.0078 | 0.139± 0.0034 | 0.143± 0.0053 | 0±0 |

**Table 4 The transition matrix between states in case of three traders. We can see that the transition frequencies between states 2 and 3 and vice versa are higher, justifying the notion of "state pair". The diagonal and anti-diagonal elements of the matrix are zero by definition.**

Checking for the stationarity of the frequency distribution of states, we study the relation $\pi T=\pi$. Table 5 shows the average of the state frequencies compared to those multiplied by the transition matrix. Though the numbers are close to each other, there is still a deviation.

| state | $\pi^t$ | $T\pi^t$ |
|---|---|---|
| 1 | 0.095 | 0.085 |
| 2 | 0.221 | 0.253 |
| 3 | 0.219 | 0.248 |
| 4 | 0.095 | 0.093 |
| 5 | 0.087 | 0.074 |
| 6 | 0.090 | 0.081 |
| 7 | 0.091 | 0.082 |
| 8 | 0.098 | 0.080 |

**Table 5 The vector of average frequencies of states and the results for the frequencies multiplied by the transition matrix. We still find deviations from stationarity.**

*5 traders*

In the case of five traders, the system has thirty-two possible states. We ran simulations for 600000 periods with strategy updating again at the end of every period. We studied the system with all the possible initial states (the codes for the states can be found in Appendix B). Figure 7 shows the mean of the frequencies of states for all possible initial states, with the standard deviations in the error bars.

In this case, having 32 possible states, we face numerical difficulties in properly sampling the strategy space and deriving a stationary distribution of frequencies, provided it exists. This can be seen from Figure 7, where the frequencies of respective states are similar, no matter the initial state; however the error bars are too large and it is impossible to check whether the frequency distribution is stationary.

Despite the noisy results, one can see that the most probable states (I1=C, I2=C, I3=F, I4=F, I5=F (state 4), I1=C, I2=F, I3=C, I4=F, I5=F (state 6), I1=F, I2=C, I3=C, I4=F, I5=F (state 7) and I1=C, I2=F, I3=F, I4=C, I5=F (state10), I1=F, I2=C, I3=F I4=C, I5=F (state 11) and I1=F, I2=F, I3=C, I4=C, I5=F (state13)) correspond to three state pairs with close frequencies (state4 – state13, state6 – state11, state7 – state10). In all these cases the most informed trader remains fundamentalist whereas the less informed traders bounce between the possible strategies similarly to the case of three traders. Once again this result supports the results shown above: It is only for the most informed trader that it is it worth using his/her information.

## Conclusions

The results of the simulations presented in our paper show a non-trivial, non-monotonic dependence of agents' returns on their amount of information, confirming results obtained in experiments (Huber et al. 2006, Kirchler and Huber 2007). We found, that average informed traders perform worse than the market level. In the simulations we analysed the case of uninformed traders and found that if there are enough traders present on the market, the uninformed, random trader is able to get the market return. Hence, we can state that average informed traders perform worse than the completely uninformed; thus in case of the average informed traders the information has a negative effect on the performance. Only the most informed traders (insiders) are able to gain above–average returns.

We analysed the return of traders in the simulations and found that the markets market globally shows informational efficiency.

In the second part of the paper we implemented a strategy updating method for traders: Periodically all traders compare their return in a certain time interval to the return of the market index in the same interval. If they find that they performed under the market they switch strategy, from fundamentalist to chartist and vice versa. We tested the equilibrium frequencies of states in the case of three and five traders, finding a set of most probable state pairs between which the system often bounces. These most probable states support the following idea on the value of information: only the most informed trader remains fundamentalist, the others switch between the strategies very often.

**Acknowledgments**
Support by OTKA T049238 is acknowledged.

## Appendix A

Below we present the MatLab® codes included in our simulation for the order putting mechanism for each of the strategies. The following notations are used in the codes: *p* stands for the price of the stock in the last transaction; *PV* stands for the *Present Value*: fundamental value of stocks conditioned on agent's information.

**Random traders**

```
if type_of_trader(trader) == 0    % random trader

rrr = rand(1);

if(rrr < .5)

ask = p + 2*randn(1);
```

```
if(ask < bestbid)

operation(trader) = -1;      %sell

else

operation(trader) = 0;

if(ask > 0)

a = [a; trader ask];
bestask = min(a(:,2));

end

end

else

bid = p + 2*randn(1);

if(bid > bestask)

operation(trader) = 1;      % buy

else

operation(trader) = 0;

if(bid > 0)

b=[b; trader bid];
bestbid=max(b(:,2));

end

end

end
```

**Fundamentalist traders**

```
elseif (type_of_trader(trader) == 1)    % fundamentalist

if(PV(trader) < bestbid)

operation(trader) = -1;         %sell

ask = PV(trader);

elseif(PV(trader) > bestask)

operation(trader) = 1;          %buy

bid = PV(trader);
```

```
else

    operation(trader) = 0;          % new bid/ask order

    distance_from_bestask = bestask - PV(trader);
    distance_from_bestbid = PV(trader) - bestbid;

    if distance_from_bestask > distance_from_bestbid

    ask = PV(trader) + (.5 * .5 * randn(1) * (PV(trader) - bestbid));

    if (ask > 0)

    a = [a; trader ask];
    bestask = min(a(:,2));

    end

    else

    bid = PV(trader) + (.5 * .5 * randn(1) * (bestask-PV(trader)));

    if (bid > 0)

    b=[b; trader bid];
    bestbid=max(b(:,2));

    end

    end

end
```

## Chartist traders

```
elseif type_of_trader(trader) == 2    % chartist

if(time > 4 & (P(time,1) < P(time-1,1) & P(time-1,1) < P(time-2,1) &
P(time-2,1) < P(time-3,1)))   % downward trend in prices

operation(trader) = -1;          %sell

ask = P(time,1) - abs(randn(1));    % ask for lower price

elseif(time > 4 & (P(time,1) > P(time-1,1) & P(time-1,1) > P(time-2,1)
& P(time-2,1) > P(time-3,1)))       % upward trend in prices

operation(trader) = 1;           %buy

bid = P(time,1) + abs(randn(1));    % offer higher price

elseif(time<4)   % random bid/ask

rd = rand(1);
```

```
if(rd<.5)

operation(trader) = -1;          %sell

ask = p - abs(randn(1));

else

operation(trader) = 1;           %buy

bid = p + abs(randn(1));

end

else                                              % no trend

operation(trader) = 0;           % new bid/ask is given as in case of
the fundamentalists with PV(trader) = p;

distance_from_bestask = bestask - p;
distance_from_bestbid = p - bestbid;

if distance_from_bestask > distance_from_bestbid

ask = p + (.5 * .5 * randn(1) * (p - bestbid));

if(ask > 0)

a = [a; trader ask];
bestask = min(a(:,2));

end

else

bid = p + (.5 * .5 * randn(1) * (bestask-p));

if(bid > 0)

b=[b; trader bid];
bestbid=max(b(:,2));

end

end

end
```

## Appendix B

Here we list the codes for the strategies used in throughout the paper. In the tables F denotes fundamentalist and C denotes chartist.

| 3 traders | | | |
|---|---|---|---|
| strategy code | I1's strategy | I2's strategy | I3's strategy |
| 1 | F | F | F |
| 2 | C | F | F |
| 3 | F | C | F |
| 4 | C | C | F |
| 5 | F | F | C |
| 6 | C | F | C |
| 7 | F | C | C |
| 8 | C | C | C |

| 5 traders | | | | | |
|---|---|---|---|---|---|
| strategy code | I1's strategy | I2's strategy | I3's strategy | I4's strategy | I5's strategy |
| 1 | F | F | F | F | F |
| 2 | C | F | F | F | F |
| 3 | F | C | F | F | F |
| 4 | C | C | F | F | F |
| 5 | F | F | C | F | F |
| 6 | C | F | C | F | F |
| 7 | F | C | C | F | F |
| 8 | C | C | C | F | F |
| 9 | F | F | F | C | F |
| 10 | C | F | F | C | F |
| 11 | F | C | F | C | F |
| 12 | C | C | F | C | F |
| 13 | F | F | C | C | F |
| 14 | C | F | C | C | F |
| 15 | F | C | C | C | F |
| 16 | C | C | C | C | F |
| 17 | F | F | F | F | C |
| 18 | C | F | F | F | C |
| 19 | F | C | F | F | C |
| 20 | C | C | F | F | C |
| 21 | F | F | C | F | C |
| 22 | C | F | C | F | C |
| 23 | F | C | C | F | C |
| 24 | C | C | C | F | C |
| 25 | F | F | F | C | C |
| 26 | C | F | F | C | C |
| 27 | F | C | F | C | C |
| 28 | C | C | F | C | C |
| 29 | F | F | C | C | C |

| 30 | C | F | C | C | C |
|----|---|---|---|---|---|
| 31 | F | C | C | C | C |
| 32 | C | C | C | C | C |

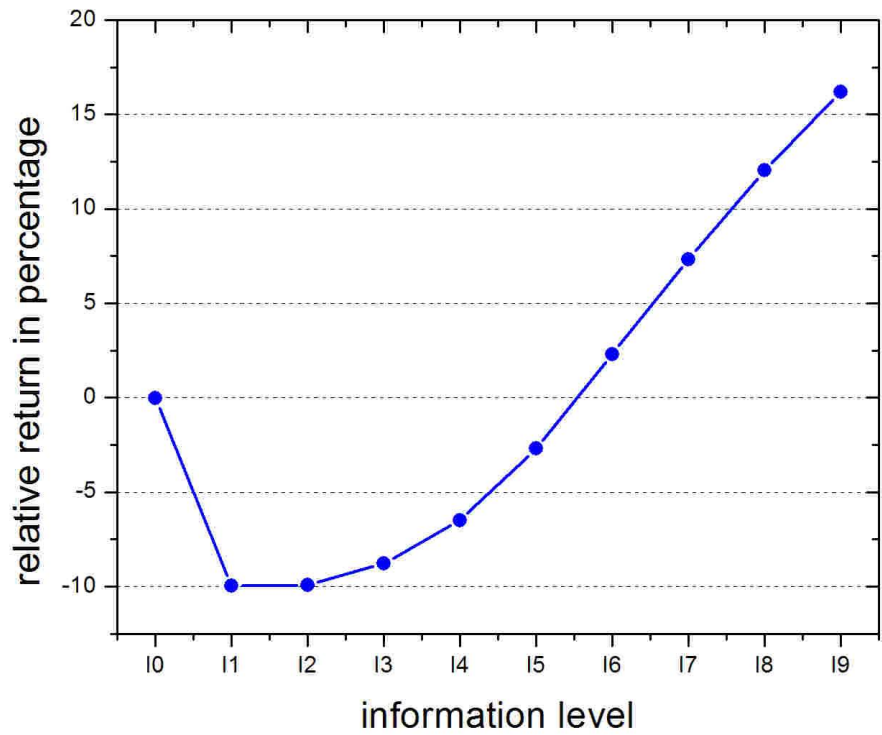

**Figure 1** Results of simulations (average of 10000 runs). Returns of traders relative to the market in percentage, as a function of information. One can see that an average level of information is not necessarily an advantage.

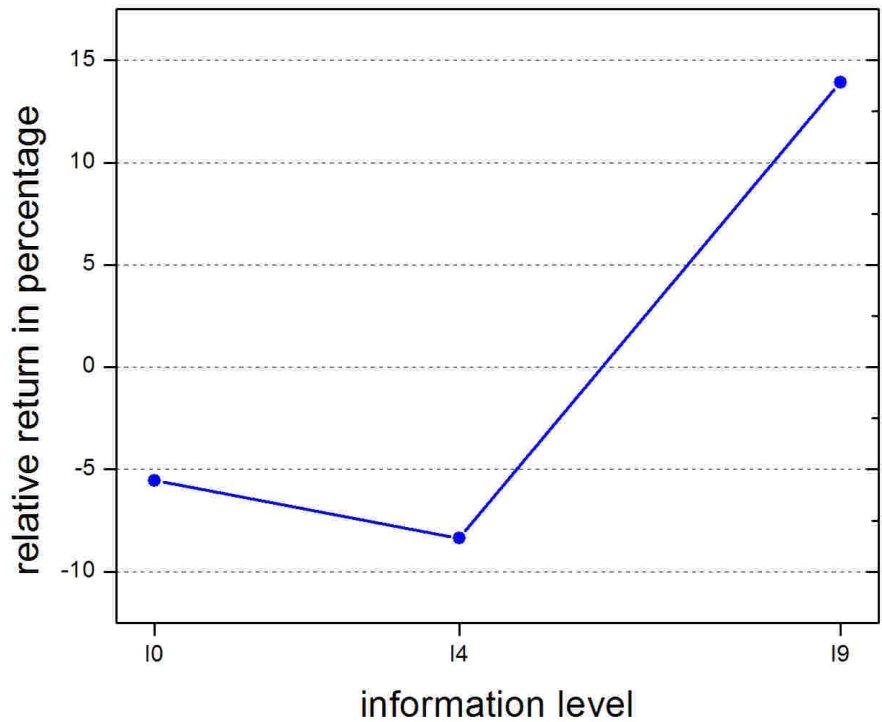

**Figure 2 Results of simulations (average of 100 runs). Returns of traders relative to the market in percentage, as a function of information. Already, in case of 3 agents one can identify a J shaped curve.**

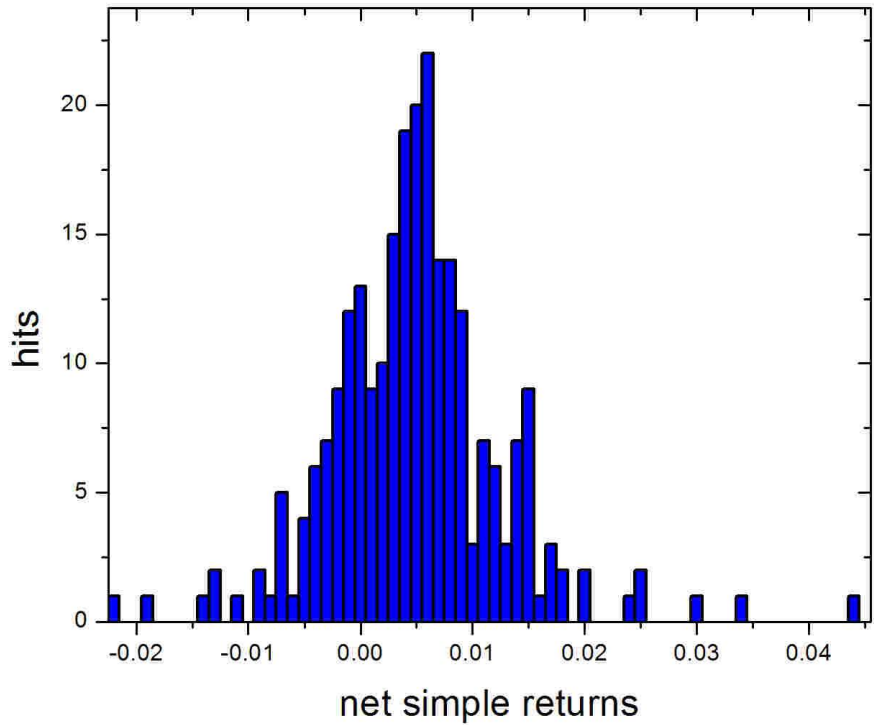

**Figure 3 The histogram of net simple returns in one session of the simulation. The conditional average of the returns is 0.0049, very near to $r_f$=0.005. Thus we can state that our market globally shows informational efficiency.**

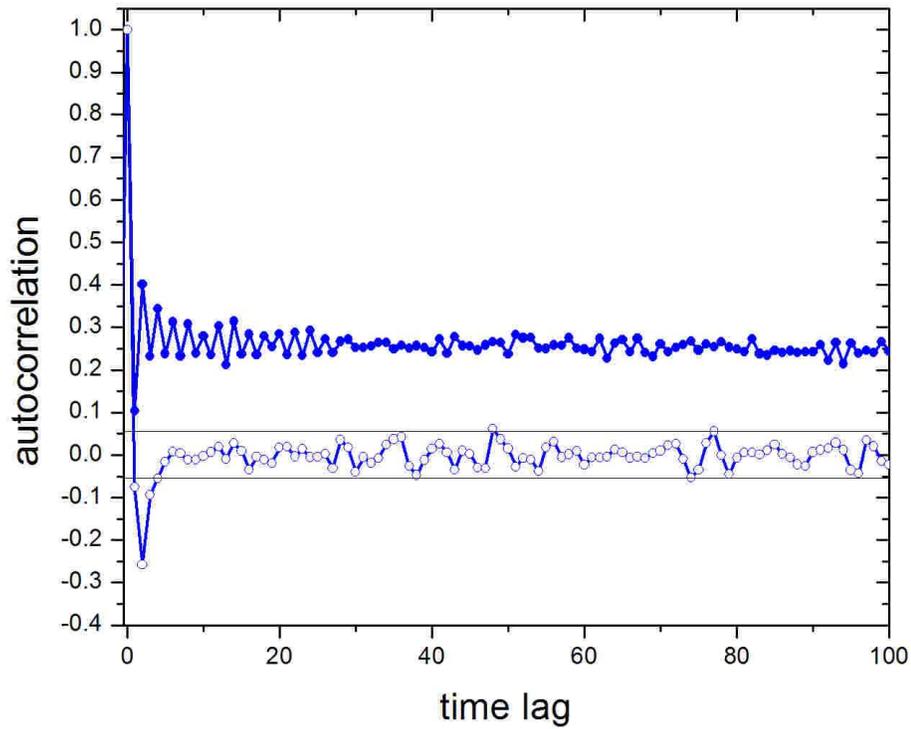

**Figure 4** Autocorrelation functions of returns (circles and lines) and absolute returns (dots and lines) and the noise level (solid lines). Autocorrelation of returns decays fast under the noise level while autocorrelation of absolute returns decays very slowly, showing the clustering of volatility. Results of one separate run of the simulations.

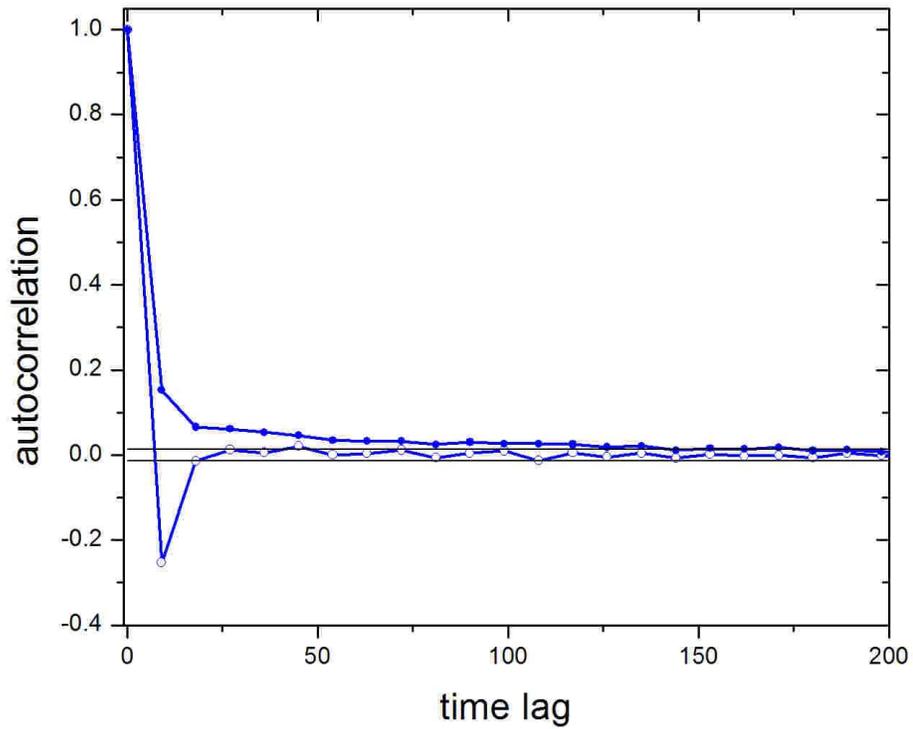

**Figure 5** Autocorrelation functions of returns (circles and lines) and absolute returns (dots and lines) and the noise level (solid lines) for the logarithmic returns of General Electric. Autocorrelation of returns decays much faster than that of the absolute returns, in agreement with the simulation results.

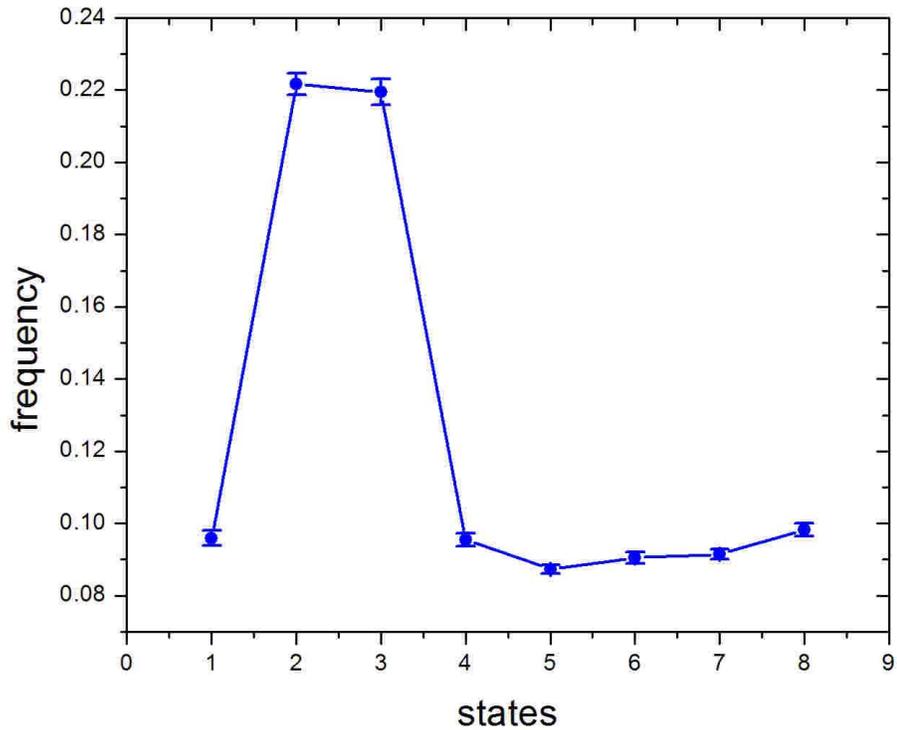

**Figure 6 The frequencies of states in case of three traders. The average over all possible initial states can be seen with the standard deviations as error bars. One can see that no matter what initial condition we start from, the frequencies of states are more or less the same. The most frequent states correspond to I1=C, I2=F, I3=F (state 2) and I1=F, I2=C, I3=F (state 3). (F=fundamentalist, C=chartist).**

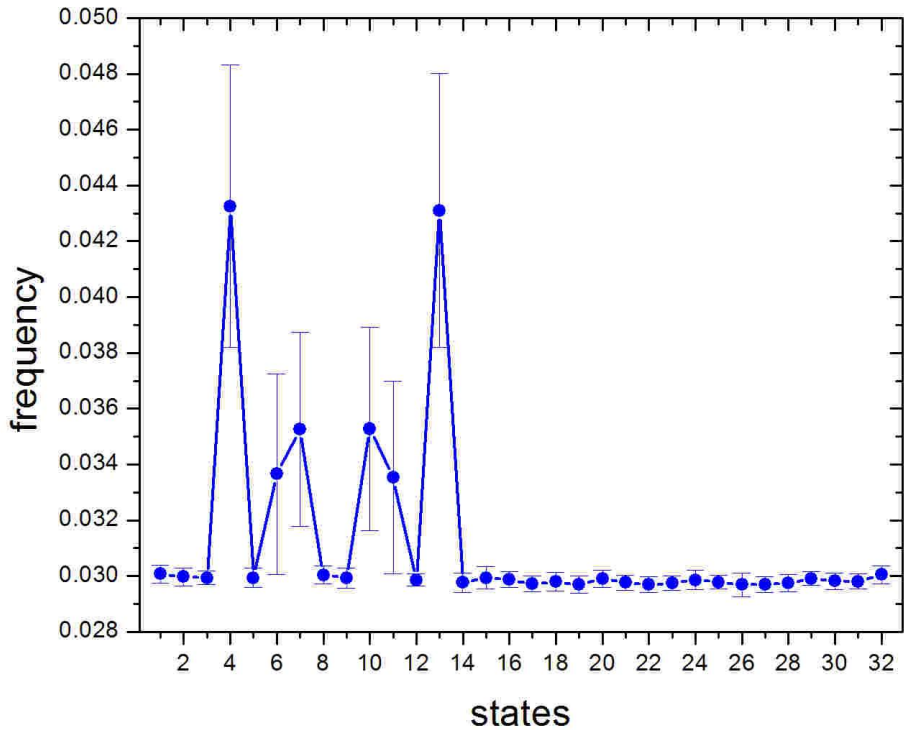

**Figure 7 The frequencies of states in case of five traders. The average over all possible initial states can be seen with the standard deviations as error bars. It can be seen that most of the states appear with low frequency regardless of the initial state. The most frequent states correspond to I1=C, I2=C, I3=F, I4=F, I5=F (state 4), I1=C, I2=F, I3=C, I4=F, I5=F (state 6), I1=F, I2=C, I3=C, I4=F, I5=F (state 7) and I1=C, I2=F, I3=F, I4=C, I5=F (state10), I1=F, I2=C, I3=F I4=C, I5=F (state 11) and I1=F, I2=F, I3=C, I4=C, I5=F (state13). (F=fundamentalist, C=chartist).**